\RequirePackage[l2tabu, orthodox]{nag}

\documentclass[conference]{IEEEtran}

\usepackage[T1]{fontenc}
\usepackage[utf8]{inputenc}
\usepackage[final,activate={true,nocompatibility},kerning=true,spacing=true]{microtype}
\usepackage[table]{xcolor}
\usepackage[pdftex,hidelinks,unicode=true]{hyperref}
\usepackage{balance}
\usepackage{paralist}
\usepackage{multirow}
\usepackage{multicol}
\usepackage{booktabs}
\usepackage{pifont}
\usepackage{float}
\usepackage{graphicx}
\usepackage{mathptmx}
\usepackage{soul}
\usepackage{array}
\usepackage{fixltx2e}

\makeatletter
\newcommand\footnoteref[1]{\protected@xdef\@thefnmark{\ref{#1}}\@footnotemark}
\makeatother

\DeclareGraphicsExtensions{.pdf, .png, .jpg}
\graphicspath{{res/}}

\newcommand{\ie}{\emph{i.e.}}
\newcommand{\eg}{\emph{e.g.}}

\newcommand{\etal}{\emph{et al.}}
\newcommand{\etc}{\emph{etc.}}
\newcommand{\adhoc}{\emph{ad hoc}}

\newcommand{\ok}{\ding{51}}
\newcommand{\notok}{\ding{55}}

\hypersetup{%
pdftitle={Classifying and Qualifying GUI Defects},
pdfauthor={Val\'eria Lelli, Arnaud Blouin, Benoit Baudry},
pdfkeywords={GUI Testing, Fault Model, Human-Computer Interaction},
bookmarksnumbered,
pdfstartview={FitH},
breaklinks=true
}

\newtheorem{definition}{Definition}

\begin{document}

\newcommand{\minitab}[2][l]{\begin{tabular}{#1}#2\end{tabular}}

\title{Classifying and Qualifying GUI Defects}

\author{
   \IEEEauthorblockN{Val\'eria Lelli}
   \IEEEauthorblockA{INSA Rennes, France\\
   \href{mailto:valeria.lelli\_leitao\_dantas@inria.fr}{valeria.lelli{\textrm\_}leitao{\textrm\_}dantas@inria.fr}}
   \and
   \IEEEauthorblockN{Arnaud Blouin}
   \IEEEauthorblockA{INSA Rennes, France\\
  \href{mailto:arnaud.blouin@irisa.fr}{arnaud.blouin@irisa.fr}}
  \and
  \IEEEauthorblockN{Benoit Baudry}
  \IEEEauthorblockA{Inria, France\\
  \href{mailto:benoit.baudry@inria.fr}{benoit.baudry@inria.fr}}
}
 
\maketitle

\begin{abstract}
% Graphical User Interface (GUI) design is currently shifting from designing GUIs composed of standard widgets (WIMP GUIs) to designing GUIs relying on more natural interactions and \adhoc{} widgets (post-WIMP GUIs).
% This shift aims at supporting the advent of GUIs providing users with more adapted and natural interactions, and the support of new input devices such as multi-touch screens.
% Standard widgets (\eg{} buttons) are replaced by \adhoc{} ones (\eg{} the drawing area of graphical editors), and interactions are shifting from mono-event (\eg{} button pressures) to multi-event interactions (\eg{} bimanual interactions).
% So, the current GUI testing approaches and their underlying fault models show their limits when applied to test such new GUIs.
Graphical user interfaces (GUIs) are integral parts of software systems that require interactions from their users. 
Software testers have paid special attention to GUI testing in the last decade, and have devised techniques that are effective in finding several kinds of GUI errors.
However, the introduction of new types of interactions in GUIs presents new kinds of errors that are not targeted by current testing techniques. %As demonstrated in this paper, however, GUIs can be affected by various kinds of errors which demand advanced testing techniques to be detected. %\nc{implying different testing techniques to be detected}.
We believe that to advance GUI testing, the community needs a comprehensive and high level GUI fault model, which incorporates all types of interactions.
%Such a fault model has to identify the faults that can affect GUIs. %to be used for evaluating the ability of GUI testing tools to find such faults.
The work detailed in this paper establishes 4 contributions:
\begin{inparaenum}
\item A GUI fault model designed to identify and classify GUI faults. 
\item An empirical analysis for assessing the relevance of the proposed fault model against failures found in real GUIs.
\item An empirical assessment of two GUI testing tools (\ie{} GUITAR and Jubula) against those failures. %failures found in real GUIs.
%\item GUI mutants we developed according to our fault model for benchmarking GUI testing tools.
\item GUI mutants we've developed according to our fault model.
These mutants are freely available and can be reused by developers for benchmarking their GUI testing tools.
\end{inparaenum}
%We show that existing GUI testing tools fail at detecting several faults and explain why to draw a research plan.
\end{abstract}

%Benoit comments
%very generic sentence: An empirical analysis for assessing the proposed fault model 
%-> can you be more specific about the type of analysis, what you look for, what is the exact purpose (more precisely than validate), what does the analysis aim at demonstrating?

%the mutants represent a contribution?
%if yes, can you explain more precisely why the mutants themselves should be considered as a contribution?

\section{Introduction}

Increasing presence of system interactivity requires software testing to closely consider the testing of graphical user interfaces (GUI). %Increasing presence instead of "The constant increase"
GUIs are composed of graphical objects called widgets, such as buttons. 
Users interact with these widgets (\eg{} press a button) to produce an action\footnote{Also called \emph{command} \cite{GAM95,BEA00b} or \emph{event} \cite{MEM07}.} that modifies the state of the system. 
For example, pressing the button "\emph{Delete}" of a drawing editor produces an action that deletes the selected shapes from the drawing. %\textit{delete}  
Most of these standard widgets provide users with an interaction composed of a single input event (\eg{} pressing a button).
In this paper we call such interactions "mono-event interactions". %For exanple instead of For instance (less formal)
These standard widgets work identically in many GUI platforms. %Removed Also to start the setence
%In this context, the GUI testing tools rely on the concept of standard widgets and
In the context of GUI testing, the tools rely on the concept of standard widgets and 
have demonstrated their ability for finding several kinds of errors in GUIs composed of such widgets, 
called WIMP\footnote{WIMP stands for \emph{Windows, Icons, Menus, and Pointing device.}} GUIs \cite{MEM07,COH12,ARL12, MAR12,NGU10}. %First time of the acronym: WIMP ]

The current trend in GUI design is the shift from designing GUIs composed of standard widgets to designing GUIs relying on more complex interactions\footnote{These interactions are more complex from a software engineering point of view. 
From a human point of view they should be more natural, \ie{} more close to how people interact with objects in the real life.} and \adhoc{} widgets \cite{BEA00b,BEL04,BLO10}. 
So, standard widgets are being more and more replaced by \adhoc{} ones. 
By \adhoc{} widgets we mean non-standard widgets developed specifically for a GUI.
Such widgets involve multi-event interactions (in opposition to mono-event interactions, \eg{} multi-touch interactions for zooming, rotating) that aim at being more adapted, natural to users. %remove also for "some widgets also..."
A simple example of such widgets is the drawing area of graphical editors with which users interact using more complex interactions such as pencil-based or multi-touch interactions.
GUIs containing such widgets are called post-WIMP GUIs \cite{DAM97}.
%As Beaudouin-Lafon wrote in 2004, "\emph{the only way to significantly improve user interfaces is to shift the research focus from designing interfaces to designing interaction}" \cite{BEL04}.
%The essential objective of this shift is the advent of GUIs providing users with more adapted and natural interactions, and the support of new input devices such as multi-touch screens.
%So, standard widgets are being more and more replaced by \adhoc{} ones. %Move this part do adhoc part
%In such a context, developers face new kinds of GUI faults that current GUI testing tools cannot detect.
The essential objective is the advent of GUIs providing users with more adapted and natural interactions, and the support of new input devices such as multi-touch screens.
As Beaudouin-Lafon wrote in 2004, "\emph{the only way to significantly improve user interfaces is to shift the research focus from designing interfaces to designing interaction}" \cite{BEL04}.
This new trend of GUI design presents to developers new problems of GUI faults that current GUI testing tools cannot detect. 
An essential pre-requisite to propose comprehensive testing techniques for both WIMP and post-WIMP GUIs is to define an exhaustive and high level GUI fault model.
Indeed, testing consists of looking for errors in a program.
%This thus requires having a clear idea about the errors we are looking for.
This requires a clear idea about the errors we are looking for.
This is the goal of fault models that permit to qualify the effectiveness of testing techniques~\cite{BOC91}. 

%In this paper, we leverage the current Human-Computer Interaction (HCI) state-of-the-art concepts and evolutions to propose an original, complete fault model for GUIs.
In this paper, we leverage of the evolution of the current Human-Computer Interaction (HCI) state-of-the-art concepts to propose an original, complete fault model for GUIs.
%This model tackles a double objective: provide a conceptual framework against which GUI testers can position their tool or technique; and % a double is used for math
This model tackles dual objectives: 
\begin{inparaenum}
\item provide a conceptual framework against which GUI testers can evaluate their tool or technique; and %position = frame?
%build a set of benchmark mutations to evaluate the ability of GUI testing tools at detecting failures for both WIMP and post-WIMP GUIs.
\item build a set of benchmark mutations to evaluate the ability of GUI testing tools to detect failures for both WIMP and post-WIMP GUIs.
\end{inparaenum}
We assess the coverage of the proposed model through an empirical analysis:
279 GUI-related bug reports of highly interactive open-source GUIs have been successfully classified using our fault model.
Also, we assess the ability of two GUI testing tools (\ie{} GUITAR and Jubula) to find real GUI failures. %Added this sentence
%Also, we selected real GUI failures classified into our fault model to assess the ability of two GUI testing tools (\ie{} GUITAR and Jubula) to detect these failures. 
Then, from an open-source system we created mutants implementing the faults described in our fault model.
These mutants are freely available and can be used for benchmarking GUI testing tools.
As an illustrative use of these mutants, we conducted an experiment to evaluate the ability of two GUI testing tools to detect these mutants. %two GUI testing tools instead of several GUI testing tools
We show that some mutants cannot be detected by current GUI testing tools and discuss future work to address the new kinds of GUI faults.

The paper is organized as follows. 
The next section examines in detail the seminal HCI concepts we leveraged to build our GUI fault model. %in detail instead of in details
Based on these concepts, the proposed GUI fault model is then detailed.
Subsequently, the benefits of our proposal are highlighted through: %Subsequently instead of "Following" to start a sentence
an empirical analysis of existing GUI bug reports; 
the manual creation of GUI mutants on an existing system; and
an evaluation of two GUI testing tools to detect such mutants.
This paper ends with related work and the conclusion presenting GUI testing challenges.

\section{Seminal HCI Concepts}\label{sec.back}

Identifying GUI faults requires an examination in detail of the major HCI concepts. %to examine in detail the major HCI concepts.
In this section we detail these concepts to highlight and explain in Section \ref{sec.faultModel} the resulting GUI faults.

Before introducing these seminal HCI concepts, we recall the basic elements that compose GUIs.
Users act on an interactive system by performing a \emph{user interaction} on a GUI.
A user interaction produces as output an \emph{action} that modifies the state of the system. 
For example, the user interaction that consists of pressing the button "\emph{Delete}" of a drawing editor produces an action that deletes the selected shapes from the drawing.
A user interaction is composed of a sequence of events (mouse move, \etc{}) produced by input devices (mouse, \etc{}) handled by users.
One interaction may involve several input devices, which is then called a multi-modal interaction.
For instance, pointing a position on a map and speaking to perform an action is a multi-modal interaction.
The correct synchronization between the different input devices is a key concern and is called multi-modal fusion.
A GUI is composed of graphical components, called widgets, laid out following a specific order.
The graphical elements displayed by a widget are either purely aesthetics (fonts, \etc{}) or presentations of data.
The state of a widget can evolve in time having effects on its graphical representation (\eg{} visibility, position, value, data content).

\emph{Direct manipulation} is one of the seminal HCI concepts \cite{SHN83,HUT85}. 
It aims at minimizing the mental effort required to use systems.
To do so, direct manipulation promotes several rules to respect while developing GUIs.
One of these rules stipulates that users have to feel engaged with control of objects of interest, not with GUIs or systems themselves.
An example of direct manipulation is the drawing area of drawing editors.
Such a drawing area represents shapes as 2D/3D graphical objects as most of the people define the concept of shapes.
Users can handle these shapes by interacting \emph{directly} within the drawing area to move or scale shapes using advanced interactions such as bi-manual interactions.
Direct manipulation is in opposition to the use of standard widgets that bring indirection between users and their objects of interest.
For instance, scaling a shape using a bi-manual interaction on its graphical representation is more direct than using a text field.
So, developing direct manipulation GUIs usually implies the development of \adhoc{} widgets, such as the drawing area.
These \adhoc{} widgets are usually more complex than standard ones since they rely on: 
advanced interactions (\eg{} bi-manual, speech+pointing interactions);
a dedicated data representation (\eg{} shapes painted in the drawing area).
Testing such heterogeneous and \adhoc{} widgets is thus a major challenge.

This contrast between GUIs composed of standard widgets only and GUIs that contain advanced widgets is reified, respectively, under the terms WIMP and post-WIMP. %\footnote{WIMP stands for \emph{Windows, Icons, Menus, and Pointing device}.} and post-WIMP.
Van Dam proposed that a post-WIMP GUI is one "\emph{containing at least one interaction technique not dependent on classical 2D widgets such as menus and icons}" \cite{DAM97}.

Another seminal HCI concept is \emph{feedback} \cite{HUT85,NOR02,BEA00b,BLO10}. 
Feedback is provided to users while they interact with GUIs.
It allows users to evaluate continuously the outcome of their interactions with the system.
Feedback is computed and provided by the system through the user interface and can take many forms.
A first simple example is when users move the cursor over a button. 
%To notify that the cursor is correctly positioned to interact with the button this last changes its shape. 
To notify that the cursor is correctly positioned to interact with the button this changes its shape.
A more sophisticated example is the selection process of most of drawing editors that can be done using a Drag-And-Drop (DnD) interaction.
While the DnD is performed on the drawing area, a temporary rectangle is painted to notify users about current selection area.

Another HCI concept is the notion of \emph{reversible actions} \cite{SHN83,HUT85,BLO10}.
The goal of reversible actions is to reduce user anxiety by about making mistakes \cite{SHN83}.
%The goal of reversible actions is to reduce user anxiety about making mistakes \cite{SHN83}. %removed this duplicated line
In WIMP GUIs, reverting actions is reified under the undo/redo features usually performed using buttons or shortcuts that revert the latest executed actions.
In post-WIMP GUIs, recent outcomes promote the ability to cancel actions in progress \cite{APP12}.

All these HCI concepts introduced in this section are interactive features that must be tested.
However, we demonstrate in this paper that current GUI fault models and GUI testing tools do not cover all these features.
In the next section, the GUI faults stemming from WIMP and post-WIMP GUIs are detailed.

\section{Fault Model}\label{sec.faultModel}

\begin{table*}[t]
\centering
\footnotesize
\caption{User Interface Faults}\label{table.GUI}
\begin{tabular}{c p{0.8cm} p{5.2cm} p{7.8cm}} 
\toprule
\textbf{Fault categories}&\textbf{ID}&\textbf{Faults} & \textbf{Possible failures}\\ \midrule
\multirow{13}*{\minitab[c]{GUI Structure \\and \\Aesthetics}} & 
GSA1
%\emph{Unexpected structure (layout or graphics) of widgets or objects}  
&Incorrect layout of widgets 

(\eg{} alignment, dimension, orientation, depth)
& The positions of 2 widgets are inverted. %\emph{WIMP GUIs -> widgets in the GUI}

A text is not fully visible since the size of text field is too small.

Rulers do not appear on the top of a drawing editor.

The vertical lines for visualizing the precise position of shapes in the drawing editor are not displayed. \\
%\Eg{} Selecting a font style in a combo box is impossible because the corresponding item is not enabled
\cmidrule(lr){2-4}
&GSA2

& Incorrect state of widgets

(\eg{} visible, activated, selected, focused, modal, editable, expandable)
& Not possible to click on a button since it is not activated.

A window is not visible so that its widgets cannot be used.

Not possible to draw in the drawing area of a drawing editor since it is not activated.\\
\cmidrule(lr){2-4}

%Post-WIMP GUIs -> objects in the GUI
&GSA3
& Incorrect appearance of widgets

(\eg{} font, color, icon, label)
& The icon of a button is not visible.

In a GUI of a power plant, the color reflecting the critical status of a pump is green instead of red.
\\
%\cmidrule(lr){2-3}

%& Wrong compound widgets

%(\eg{} check box, radio button, toggle button)

%& The state of a toggle button does not correspond with its label. \\
%\cmidrule(lr){2-3}

%& Wrong appearance of objects or grouping objects

%(\eg{} compose an image or template)

%& The texture of the background image is blurry. \\
%\cmidrule(lr){2-3}

%& Incorrect properties of I/O widgets

%(\eg{} size, data type)
%& Object's handlers are not contrasted with the GUI background color.\\
\midrule

\multirow{8}*{\minitab[c]{Data \\ Presentation}} %& 
%\emph{Displaying unexpected data}
%& \textit{WIMP GUIs -> simple value data}\\
%&Incorrect Data layout

%(\eg{} alignment: justified or left, orientation, dimension, depth) \comment{Similar to widgets but with some particularities}
%& The text displayed on a text box is cut.\\
%\cmidrule(lr){2-3}

%&Incorrect Data appearance

%(\eg{} font, color, style: bold, italic, texture, shadow, opacity, border) \comment{Similar to widgets but with some particularities}
%& Drawing a shape (by drag and drop) on canvas displayed it with a wrong border \\
%\cmidrule(lr){2-3}

%The rendering concept is only applied to objects: complex data
&DT1

& Incorrect data rendering

(\eg{} scaling factors, rotating, converting) 
%or \emph{wrong appearance} (\eg{} color, texture, opacity, shadow), or \emph{wrong layout} (\eg{} position, size, handlers, border, geometry: angle, depth) 
& The size of a text is not scaled properly.

In a drawing editor, a dotted line is painted as a dashed one.

A rectangle is painted as an ellipse. \\

%Drawing a shape on canvas displayed it with a wrong border. \\ It is similar to third example.
\cmidrule(lr){2-4}
&DT2

& Incorrect data properties

%Hyperlink/Hypertext is focused
(\eg{} selectable, focused)
& A web address in a text is not displayed as hyperlink. \\

% It is the GUI structure fault -> incorrect state of widgets
%\fixme{Texts are overlapping each other in a text area.}

%\fixme{A shape on a drawing editor is not editable.} \\
\cmidrule(lr){2-4} 
&DT3

& Incorrect data type or format

(\eg{} degree \emph{vs} radian, float \emph{vs} double)
%\emph{simple data} (\eg{} string, float, degree, pixel, inch, date, time), or  \emph{complex data} (jpg, svg, bitmap, pdf, avi)
%&A text field to save a file only accepts numerical values.
&The date is displayed with five digits (\eg{} dd/mm/y) instead of 6 digits (\eg{} dd/mm/yy). 

A text field displays an angle in radian instead of in degree. \\

%An PNG image is not displayed in a drawing editor.\\
%\cmidrule(lr){2-3}

%This fault is into incorrect data rendering
%& Incorrect animation 

%(\eg{} direction, delay, duration) 
%& The animation resulting from scrolling a long document does not work properly.\\
\midrule
\end{tabular}
%\caption{User Interface Faults}\label{table.GUI}
\end{table*}

%--------------------------------------

\begin {table*}
\centering
\small
\caption{User Interaction Faults}\label{table.UI}
\begin{tabular}{c p{0.8cm}p{5.4cm}p{7.8cm}} 
\toprule 
\textbf{Fault categories} & \textbf{ID} & \textbf{Faults} & \textbf{Possible failures}\\
\midrule

\multirow{4}*{\minitab[c]{Interaction \\ Behavior}}
& IB1
&  Incorrect behavior of a user interaction
& A bi-manual interaction developed for a specific purpose does not work properly.

The synchronization between the voice and the gesture does not work properly in a voice+gesture interaction.\\
%\cmidrule(lr){2-3}
% This fault is part of the incorrect action results.
%&  Incorrect handling of I/O operations.
%& Typing a large number on a text field crashes the GUI.\\
\midrule

\multirow{8}*{\minitab[c]{Action}}
&ACT1 
& Incorrect action results
& Translating a shape to a position $(x,y)$ translates it to the position $(-x,-y)$.

Setting the zoom level at 150\%, sets it at 50\%.\\

%\fixme{Saving a picture as a PNG saves it as a JPG.}\\%It is not a GUI fault!
\cmidrule(lr){2-4}

&ACT2
& No action executed
& Clicking on a button has no effect.

Executing a DnD on a drawing area to draw a rectangle has no effect.\\
\cmidrule(lr){2-4}

&ACT1 
& Incorrect action executed
& Clicking on the button \emph{Save} shows the dialogue box used for loading.

Scaling a shape results in its rotation.

Performing a DnD to translate shapes results in their selection.
\\ 
\midrule

% \multirow{8}*{\minitab[c]{Interaction \\ Effects}}
% & User interaction does not activate dependent actions 
% & Clicking on a "play" button does not change it for "pause"\\
% \cmidrule(lr){2-3} 
% 
% & Do not activate correctly dependent actions through widgets events 
% &\\
% \cmidrule(lr){2-3} 
% 
% & Do not activate correctly dependent actions through objects interactions 
% & Selecting a shape on canvas does not enable the button "delete" on a toolbar. Or, selecting a shape does not enable its handler "resize" or "rotate"\\
% \midrule

\multirow{8}*{\minitab[c]{Reversibility}} 
&RVSB1
& Incorrect results of undo or redo operations
& Clicking on the button \emph{redo} does not re-apply the latest undone action as expected.

Pressing the keys \emph{ctrl+z} does not revert the latest executed action as expected.\\
\cmidrule(lr){2-4}

&RVSB2
& Reverting the current interaction in progress works incorrectly
& Pressing the key "\emph{Escape}" during a DnD does not abort this last.

Saying the word "\emph{Stop}" does not stop the interaction in progress.\\
\cmidrule(lr){2-4}

&RVSB3
& Reverting the current action in progress works incorrectly
& Clicking on the button "\emph{Cancel}" to stop the loading of the file previously selected does not work properly.\\

% & User interaction cannot be aborted and its ongoing actions & \\
% & Simple Interactions: actions triggered \textit{after} an interaction cannot be aborted: pressing a button or a key to stop the ongoing action does not work
% & Clicking on a cancel button to stop loading an image does not interrupt this action\\ 
% & Complex Interactions:  actions triggered \textit{while} an interaction is being done cannot be aborted: 
% releasing the mouse button or fingers or pressing keys while manipulated an object does not abort the ongoing action & \\
% & & Pressing a key (\eg{} escape) while a shape is being drawn (by DnD) on canvas does not abort the drawing.\\
\midrule

\multirow{5}*{\minitab[c]{Feedback}}
&FDBK1
& Feedback provided by widgets to reflect the current state of an action in progress works incorrectly.
%Showing incorrectly a feedback by widgets after a user interaction: widgets are not highlighted to show they are selected; missing or wrong a haptic feedback on touch screens; or no widgets to show actions in progress or show it incorrectly, so on
& The progress bar that shows the loading progress of a file works incorrectly.\\
%Do not show up a toolbar progress (or show an incorrect one) while a package is being installed.
\cmidrule(lr){2-4}

&FDBK2
& The temporary feedback provided all along the execution of long interactions is incorrect.
%Showing incorrectly the intermediary states of the ongoing action during a user interaction: do not show the discrete actions that compose an complex interaction 
& Given a drawing editor, drawing a rectangle using a DnD interaction does not show the created rectangle during the DnD as expected.
\\
\midrule

% \multirow{8}*{\minitab[c]{Reversible \\ Actions}} & Actions over GUIs cannot be reversible as expected & \textit{WIMP GUIs}\\
% & Simple Interactions: actions performed over widgets \textit{after} an interaction are not reverted properly &
% The action "undo color text" cannot be undone\\ 
% & Complex Interactions:  \FIXME{discrete} actions performed over objects \textit{during} an interaction are not reverted properly & \textit{Post-WIMP GUIs} \\
% & & The action "create a curve" is undone like a single one instead of reverting (step by step) its discrete actions performed to create that curve\\
% \bottomrule
\end{tabular}
%\caption{User Interaction Faults}\label{table.UI}
\end{table*}

In this section we present an exhaustive GUI fault model.

Bochmann \etal{} \cite{BOC91} define a fault model as:

\begin{definition}[Fault Model]
 A fault model describes a set of faults responsible for a failure possibly at a higher level of abstraction. %and is the basis for mutation testing. %\cite{BOC91}.{According to Bochmann}
\end{definition}

%\noindent That definition helps clarify what a fault is:
\noindent To recall what a fault is:
\begin{definition}[Fault]
Faults are textual (or graphical) differences between an incorrect and a correct behavior description \cite{PRE13}.
\end{definition}

Based on these definitions, %testing GUIs consists of verifying whether the GUI of the system under test (SUT) has the expected behavior, \ie{} detecting GUI failures.
we propose the following definitions of a GUI fault and failure:

%BB comment: do we really need theses 3 definitions? I find them pretty similar to each other . Can't we just have definition 2?
\begin{definition}[GUI Fault]
GUI faults are differences between an incorrect and a correct behavior description of a GUI.
\end{definition}

\begin{definition}[GUI Error]
A GUI error is an activation of a GUI fault that leads to an unexpected GUI state. 
\end{definition}

\begin{definition}[GUI Failure]
A GUI failure is a manifestation of an unexpected GUI state provoked by a GUI fault. 
\end{definition}

A GUI fault can be introduced at different levels of a GUI software (\eg{} GUI code, GUI models).
An illustration of a GUI fault is: a correct line of GUI code \emph{vs} an incorrect line of GUI code.
%Is it the entry/input that is unexpected/wrong (precondition violation) or is it the code that does not handle the input correctly (a bug): ???
For example, a GUI fault can be activated when an \emph{unexpected entry}, such as a wrong value into an input widget, is not \emph{handled correctly} by its GUI code. %its corresponding GUI code.
%So, an unexpected GUI state, resulting \emph{for instance is a crash}, is manifested when a \emph{user clicks on a button} after typing that entry.
So, an unexpected GUI state is manifested (\eg{} a crash as a GUI failure)  when a user clicks on a button after typing this entry.

%In this section we present an exhaustive GUI fault model.
%To build the fault model we first analyzed the state-of-the-art of HCI concepts (see Section \ref{sec.back}).
%%We then analyzed real GUI bug reports (different than those used in Section~\ref{sec.eval1}) to assess and precise the fault model.
%We then analyzed real GUI bug reports (different than those used in Section~\ref{sec.eval1}) to assess and to precise the fault model.
%We performed a round trip process (\ie{} \FIXME{incremental iterative feedback process}) between the analysis of HCI concepts and GUI bug reports until a stable fault model was obtained.%instead of is obtained
To build the proposed fault GUI model we first analyzed the state-of-the-art of HCI concepts (see Section \ref{sec.back}).
We then analyzed real GUI bug reports (different than those used in Section~\ref{sec.eval1}) to assess and to precise the fault model.
We performed a round trip process between the analysis of HCI concepts and GUI bug reports until obtain a stable fault model. %(\ie{} \FIXME{incremental iterative feedback process?}): need space

The description of our fault model is divided into two groups: 
The user interface faults and the user interaction faults. 
The user interface faults refer to faults affecting the structure and the behavior of graphical components of GUIs.
The user interaction faults refer to faults affecting the interaction process when a user interacts with a GUI.

\subsection{User Interface Faults}

GUIs are composed of widgets that can act as mediators to interact indirectly (\eg{} buttons on WIMP GUIs) or directly (direct manipulation principle in post-WIMP GUIs) with objects of the data model.
In this section, we categorize the user interface faults, \ie{} faults related to the structure, the behavior, and the appearance of GUIs.
%We identified 2 categories of user interface faults:
We further break down user interface into two categories:
the \emph{GUI structure and aesthetics}, and the \emph{data presentation} fault, as introduced below. %aesthetics instead of aestheticism
Table \ref{table.GUI} presents an overview of these faults and their potential failures.

\subsubsection{GUI Structure and Aesthetics Fault}

This fault category corresponds to unexpected GUI designs.
Since GUIs are composed of widgets laid out following a given order, the first fault is the \emph{incorrect layout of widgets} (GSA1).
Possible failures corresponding to this fault occur when GUI widgets follow an unexpected layout (\eg{} wrong size or position).
The next fault concerns the \emph{incorrect state of widgets} (GSA2).
Widgets' behavior is dynamic and can be in different states such as visible, enabled, or selected.
This fault occurs when the current state of a widget differs from the expected one.
For example, a widget is unexpectedly visible.
The following fault treats the \emph{unexpected appearance of widgets} (GSA3).
That concerns aesthetic aspects of widgets not bound to the data model, such as look-and-feels, fonts, icons, or misspellings. %aesthetics

\subsubsection{Data presentation}

In many cases, widgets aim at editing and visualizing data of the data model.
For example with WIMP GUIs, text fields or lists can display simple data to be edited by users.
Post-WIMP GUIs share this same principle with the difference that the data representation is usually \adhoc{} and more complex.
For example, the drawing area of a drawing editor paints shapes of the data model.
Such a drawing area has been developed for the specific case of this editor.
That permits to represent graphically in a single widget complex data (\eg{} shapes).
In other cases, widgets aim at monitoring data only.
This is notably the case for some GUIs in control commands of power plants where data are not edited but monitored by users. %bb comment: very specific example but probably not the best one to illustrate the concept.
The definition of data representations is complex and error-prone.
It thus requires adequate \emph{data presentation} faults.

The first fault of this category is the \emph{incorrect data rendering} (DT1). %\comment{In italic but the next faults it is no more the case}
DT1 is provoked when data is converted or scaled wrongly. 
Possible failures for this fault are manifested by unexpected data appearance (\eg{} wrong color, texture, opacity, shadow) or data layout (\eg{} wrong position, geometry).
The second fault concerns incorrect data properties (DT2).
Properties define specific visualization of data such as selectable or focused.
A possible failure is a web address that is not displayed as a hyperlink.
The last fault (DT3) occurs when an incorrect data type or format is displayed.
For instance, an angle value is displayed in radian instead of in degree.

\subsection{User Interaction Faults}
In this section, we introduce the faults that treat user interactions.
The proposed faults are based on the characteristics of WIMP and post-WIMP GUIs detailed in the previous section.
For each fault we separated our analysis into two parts.
One dedicated to WIMP interactions and another one to post-WIMP interactions.
WIMP interactions refer to interactions performed on WIMP widgets.
They are simple and composed of few events (click\footnote{A click is one interaction composed of the event \emph{mouse pressed} followed by the event \emph{mouse released}. Its simple behavior has leaded to consider a click as an event itself.}, key pressed, \etc{}).
Post-WIMP interactions refer to interactions performed on post-WIMP widgets.
Such interactions are more complex since they can
be multimodal, \ie involve multiple input devices (gesture, gyroscope, multi-touch screen);
be concurrent (\eg{} in bi-manual interactions the two hands evolve in parallel);
be composed of numerous events (\eg{} multimodal interactions may be composed of sequences of pressure, move, and voice events).
%be composed of numerous events (\eg{} the DnD interaction is composed of an ordered sequence of mouse press, move, and release events).
Such interactions can be modeled as finite-state machines \cite{BLO10,BLO11,APP08}.
Subsequently the direct manipulation principles, other particularities of post-WIMP interactions are that they aim at: %Subsequently instead of Following to start a sentece
being as natural as possible;
providing users with the feeling of handling data directly (\eg{} shapes in drawing editors).
Table \ref{table.UI} summarizes the \emph{user interaction} faults and some of their potential failures for both WIMP and post-WIMP interactions. 
These faults are detailed as follows.

\subsubsection{Interaction Behavior}

Developing post-WIMP interactions is complex and error-prone.
Indeed, as explained in the section on GUIs' characteristics, it may involve many sequences of events or require the fusion of several modalities such as voice and gesture.
So, the first fault (IB1) occurs when the behavior of the developed interactions does not work properly.
This fault mainly concerns post-WIMP widgets since WIMP widgets embed simple and hard-coded interactions.
For instance, an event such as \emph{pressure} can be missing in a bi-manual interaction.
Another example is the incorrect synchronization between the voice and the gesture in a voice+gesture interaction.

\subsubsection{Action}

%This category of faults regroups faults that concern actions produced while interacting with the system.
This category of faults groups faults that concern actions produced while interacting with the system.
The first fault (ACT1) focuses on the incorrect results of actions.
In this case the expected action is executed but its results are not correct.
For instance with a drawing editor, a failure can be the translation of one shape to the given position $(-x,-y)$ while the position $(x,y)$ was expected.
%The origin of this failure can be located in the action itself or in its settings.
The root cause of this failure can be located in the action itself or in its settings.
For instance, a first root cause of the previous failure can be the incorrect coding of the translation operation. 
A second root cause can be located in the settings of the translation action.

The second fault (ACT2) concerns the absence of action when interacting with the system.
For instance, this fault can occur when an interaction, such as a keyboard shortcut, is not correctly bound to its widget.

The third fault (ACT3) consists of the execution of wrong actions.
The root cause of this fault can be that the wrong action is bound to a widget at a given instant.
For instance: 
clicking on the button \emph{Save} shows the dialogue box used for loading;
doing a DnD interaction on a drawing area selects shapes instead of translating them.

\subsubsection{Reversibility}

%This fault category regroups 3 faults.
This fault category groups three faults.
The first fault (RVSB1) concerns the incorrect behavior of the undo/redo operations.
Undo and redo operations usually rely on WIMP widgets such as buttons and key shortcuts.
These operations revert or re-execute actions \emph{already terminated} and stored by the system.
A possible failure is the incorrect reversion of the latest executed action when the key shortcut \emph{ctrl+z} is used.

Contrary to WIMP interactions, that are mainly one-shot, many interactions last some time such as the DnD interaction.
In such a case, users may be able to stop an interaction in progress.
The second fault (RVSB2) thus consists of the incorrect interruption of the current interaction in progress.
For instance, pressing the key "\emph{Escape}" during a DnD does not stop this last.
This fault could have been classified as an interaction behavior fault.
We decided to consider it as a reversibility fault since it concerns the ability to revert an ongoing interaction.

Once launched, actions may take time to be executed entirely.
In this case such actions can be interrupted.
The third fault (RVSB3) concerns the incorrect interruption of an action in progress.
A possible failure concerns the file loading operation:
clicking on the button "\emph{Cancel}" to stop the loading of a file does not work properly.

\subsubsection{Feedback}

Widgets are designed to provide immediate and continuous feedback to users while they interact with them.
For instance, progress bars showing the loading progress of a file is a kind of feedback provided to users.
The first fault of this category (FDBK1) concerns the incorrect feedback provided by widgets to reflect the current state of an action in progress.
This fault focuses on actions that last in time and \emph{which progress should be monitored} by users.

The second fault (FDBK2) focuses on potentially long interactions (\ie{} interactions taking a certain amount of time to be completed) which progress should be discernible by users.
For instance with a drawing editor, when drawing a shape on the drawing area, the shape in creation should be visible so that the user knows the progression of her work.
So, a possible failure is drawing a rectangle using a DnD interaction, that works correctly, does not show the created rectangle during the DnD as expected.

\subsection{Discussion}

The definition and the use of a fault model raise several questions we discuss about in this sub-section.

\emph{What are the benefits of the proposed GUI fault model?}
The benefits of our GUI fault model are twofold.
First, a fault model is an exhaustive classification of faults for a specific concern \cite{BOC91}.
Providing a GUI fault model permits GUI developers and testers to have a precise idea of the different faults they must consider.
%Yet, our fault model has been defined on the basis of the major HCI concepts, introduced in Section \ref{sec.back}, covering both WIMP and post-WIMP GUIs. %pass 10 pages
As an illustration, Section \ref{sec.eval1} describes an empirical analysis we conducted to classify and discuss about GUI failures of open-source GUIs.
Second, our GUI fault model allows developers of GUI testing tools to evaluate the efficiency of their tool in terms of bug detection power w.r.t. a GUI specific fault model.
As detailed in Section \ref{sec.eval2}, we created mutants of an existing GUI. %BB comment: what GUI? how are the mutants distributed / made available to the developers of testing tools? We have provided this information on section VI
Each mutant contains one GUI failure that corresponds to one GUI fault of our fault model.
Developers of GUI testing tools can run their tools against these mutants for benchmarking purposes.

\emph{Should usability have been a GUI fault?} 
Answering this question requires the definition of a fault to be re-explained:
a fault is a difference between the observed behavior description and the expected one.
Usability issues consist of reporting that the current observed behavior of a specific part of a GUI lacks at being somehow usable.
That does not mean the observed behavior differs from the behavior expected by test oracles.
Instead, it usually means that the expected behavior has not been defined correctly regarding some usability criteria.
That is why we do not consider usability as a GUI fault. This reasoning can be extended to other concerns such as performance.

%BB comment: if this section is about the classification of faults, why have two questions about the classification of failures?
\emph{How to classify GUI failures into a fault model?} 
A GUI failure is a perceivable manifestation of a GUI error.
Classifying GUI failures thus requires to have identified the root cause (\ie{} GUI error) of the failure.
So, classifying GUI failures can be done by experts of the GUI under test.
%These experts need sufficient information, such as patches, logs, or stack traces, to identify the root cause of the failure to classify it.
These experts need sufficient information, such as patches, logs, or stack traces, to identify if the root cause of a failure is a GUI error to then classify it.
For example, given a failure manifested in the GUI and caused by a precondition violation. %occurs when a precondition is violated. is caused by a precondition violation
%For example, a root cause of a failure manisfested in a GUI can be a precondition violation. 
In this case, such a failure is not classified into the GUI fault model.
Similarly, classifying correctly a GUI failure also requires to qualify the involved widgets (\eg{} standard or \adhoc) as well as
the interaction (\eg{} mono-event or multiple-event interaction). % (\ie{} action) or multiple-event interaction (\ie{} interaction).

\emph{How to classify failures stemming from other failures?}
For instance, the incorrect results of the execution of an action (action fault) let a widget not visible as expected (GUI structure fault).
In such cases, only the first failure must be considered since it puts the GUI in an unexpected and possibly unstable state.
Besides, the appearance of a GUI error depends on the previous actions and interactions successfully executed.
Typical examples are the undo and redo actions.
A redo action can be executed only if an action has been previously performed.
Furthermore, the success of a redo action may depend on the previous executed actions.
We considered this point during the creation of mutants (as detailed in Section \ref{sec.eval2}) to provide failures that appear both with and without previous actions.

%Need space!
%In the next 3 sections, we present empirical studies that assess the relevance of our GUI fault model when applied to real situations.

\section{Relevance of the Fault Model:\\ an empirical analysis}\label{sec.eval1}

\begin {table*}[th]
\scriptsize\setlength{\tabcolsep}{0.12cm}
\caption{Distribution of analyzed failures per software}\label{appfailures}
\centering
\small
\begin{tabular}{lcccl} 
\hline 
\textbf{Software} & \textbf{Analyzed failures} & \textbf{User interface failures} & \textbf{User interaction failures } & \textbf{Repositories link}\\ \hline
Sweet Home 3D&33&55\%&45\%&\url{http://sourceforge.net/p/sweethome3d/bugs/}\\ 
File-roller&32&28\%&72\%&\url{https://bugzilla.gnome.org/query.cgi}\\
JabRef&84&42\%&58\%&\url{http://sourceforge.net/p/jabref/bugs/}\\ 
Inkscape &82&28\% &72\%&\url{https://bugs.launchpad.net/inkscape/}\\ 
Firefox Android&48&60\%&40\%&\url{https://bugzilla.mozilla.org/}\\ \hline
\end{tabular}
%\caption{Distribution of analyzed failures per software}\label{appfailures}
\end{table*}

In this section the proposed GUI fault model is evaluated.
Our evaluation has been conducted by an empirical analysis to assess the relevance of the model w.r.t. faults currently observed in existing GUIs.
The goal is to state whether our GUI fault model is relevant against failures found in real GUIs.

\subsection{Introduction}

To assess the proposed fault model, we analyzed bug reports of 5 popular open-source software systems:
Sweet Home 3D, File-roller, JabRef, Inkscape, and Firefox Android.
These systems implement various kinds of widgets, interactions, and encompass different platforms (desktop and mobile).
%Their GUI features cover both the indirect and direct manipulations.
%Thus, complex interactions are provided such as drag and drop.
%The interactions are enabled by using different input devices (\eg{} keyboard: shortcuts to execute an action).
%Besides, they allow to manipulate discrete data (\eg{} vector-based graphics), which actions can be undone and redone.
Their GUIs cover the main following features:  
\emph{indirect and direct} manipulation;
\emph{several input devices} (\eg{} mouse, keyboard, touch);
\adhoc{} \emph{widgets} such as canvas;
\emph{discrete data manipulation} (\eg{} vector-based graphics); and
\emph{undo/redo} actions.

\subsection{Experimental Protocol}
%Bug reports have been selected manually from the researcher/tester perspective by analyzing only data available in the failures report (\ie{} black box analysis).
Bug reports have been analyzed manually from the researcher/tester perspective by looking only at data available in the failures report (\ie{} black box analysis).
To focus on detailed and commented bug reports that concern GUI failures, the selection has been driven by the following rules.
Only closed, fixed, and in progress bug reports were selected.
The following search string has been also used to reduce the resulting sample:
\emph{interface OR "user interface” OR “graphical user interface” OR "graphical interface" OR GUI OR UI OR layout OR design OR graphic OR interaction OR “user interaction” OR interact OR action OR feedback OR revert OR reversible OR undo OR redo OR abort OR stop OR cancel}.
Each report has been then manually analyzed to state whether it is a GUI failure.
Also, selected bug reports have to provide explanations about the root cause of the failure such as a patch or comments.
This step is crucial to be able to categorize the failures using our GUI fault model considering their root cause.
We also discarded failures identified as non-reproducible, duplicated, usability, or user misunderstanding.
From this selection we kept 279 bug reports (in total for the five systems) describing one GUI failure each. %279 out of how many in the bug reports?
The following sub-sections discuss about these failures and the classification process.

\subsection{Classification and Analysis}

\begin{figure}[H]
\centering
	\includegraphics[scale=0.535]{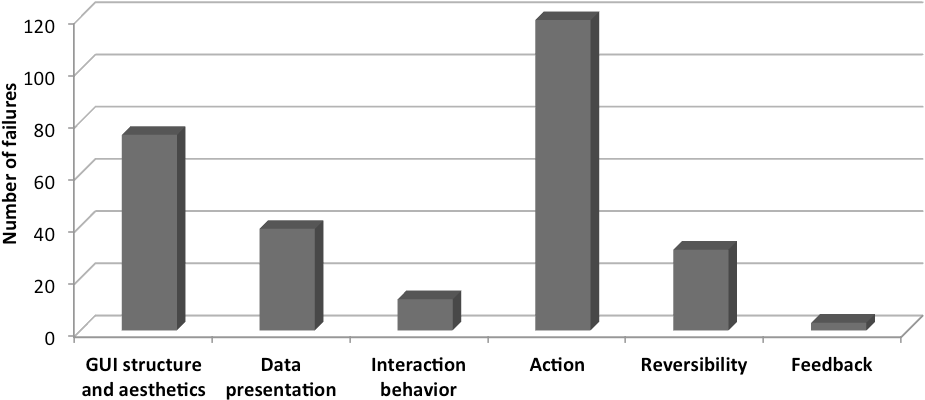}
	\caption{Classification of the 279 bug reports using the GUI fault model} \label{failuresFM}
\end{figure}

All the 279 failures have been successfully classified into our fault model.
Fig. \ref{failuresFM} gives an overview of the selected bug reports classified using our proposed fault model.
These failures were classified into the \emph{Action} (119 failures, 43\%),
\emph{GUI Structure and Aesthetics} (75 failures, 27\%), \emph{Data Presentation} (39 failures, 14\%), \emph{Reversibility} (31 failures, 11\%), \emph{Interaction behavior} (12 failures, 4\%), and \emph{Feedback} (3 failures, 1\%) fault categories.
Most of the failures classified into \emph{GUI Structure and Aesthetics} concern the \emph {incorrect layout of widgets} (51\%).
Likewise, most of the failures in the \emph{Action} category refer to \emph {incorrect action results} (75\%).

Table \ref{appfailures} shows the distribution of the 279 analyzed GUI failures per software and category (user interface or user interaction).
These results point out that the systems \emph{Sweet Home 3D} and \emph{Firefox Android} seem to be more affected by user interface failures. 
Most of these failures concern the \emph{GUI structure and aesthetics} fault. 
That can be explained by the complex and \adhoc{} GUI structure of these systems.
\emph{File Roller} and \emph{JabRef} GUIs include widgets with coarse-grained properties (\ie{} simple input value such as number or text).
Most of their failures concern WIMP interactions classified into the \emph{action} category.
In contrast, \emph{Inkscape} presented more failures classified as post-WIMP.
Indeed, Inkscape, a vector graphics software, mainly relies on its drawing area that provides users with different post-WIMP interactions.
These failures have been categorized mainly into \emph{interaction behavior}, \emph{action}, and \emph{reversibility}.

\begin{figure}[H]
\centering
	\includegraphics[scale=0.65]{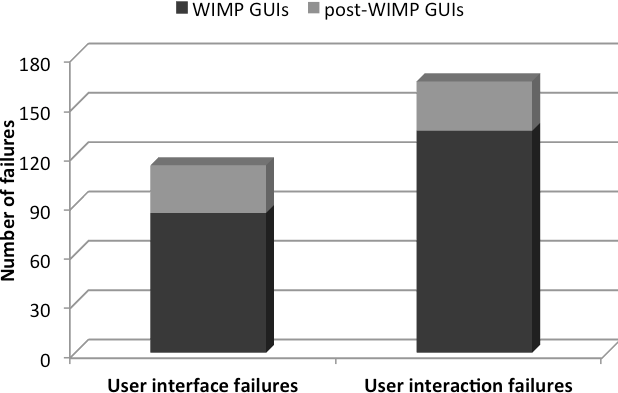}
	\caption{Manifestation of failures in the user interface and user interaction levels} \label{failuresGUI}
\end{figure}

As depicted by Fig. \ref{failuresGUI}, 41\% of these 279 GUI failures are originated by faults classified into the user interface category and 59\% into the user interaction category.
Most of user interaction failures have been classified into the \emph{incorrect action results} (54\%).
This plot also highlights that only 25\% of the analyzed user interface failures and 18\% of the user interaction ones have been classified as post-WIMP.
We comment these results in the following sub-section.

\subsection{Discussion}
%BB coment: which results? the fact that there are much more WIMP failures than post-WIMP? 
%These results%
The empirical results must be balanced with the fact that user interactions are less tangible than user interfaces.
So, users may report more GUI failures when they can perceive failures graphically (an issue in the layout of a GUI or in the result of an action visible through a GUI).
Users, however, may have difficulties to detect a failure in an interaction itself while interacting with the GUI.
That may explain the low number of failures (4\%) classified into \emph{Interaction Behavior}.
Another explanation may be the primary use of WIMP widgets, relying on simple interactions.

In our analysis, many failures that could be related to \emph{Feedback} were discarded since they concerned enhancements or usability issues, which are out of the scope of a GUI fault model as discussed previously.
For instance, GUI failures that concern the lack of haptic feedback in Firefox Android were discarded.
So, few faults (1\%) were classified into this category.
Another explanation may be the difficulty for users to identify feedback issues as real failures that should be reported.

%% FIXME what does it mean? how many failures are concerned? The following paragraph is not clear.
We observed that some reported GUI failures are false positives regarding the \emph{fault localization}: %\cite{GON12}:
if the report does not have enough information about the root cause of a failure (\eg{} patch or exception log), a GUI failure can be classified in a wrong fault category.
For example, when moving a shape using a DnD does not move it. 
At a first glance, the root cause of this failure can be associated to an incorrect behavior of the DnD.
So, this failure can be categorized into the interaction behavior. 
However, after analyzing the root cause this failure refers to an action failure since the DnD works properly, but no action is linked to this interaction.

Likewise, the failures related to \emph{Reversibility} and \emph{Feedback} were easily identified through the steps to reproduce them.
For example in JabRef, "\emph{pressing the button "Undo" will clear all the text in the field, but then pressing the button "Redo" will not recover the text}".
Furthermore, some systems do not revert interactions step by step but entirely.
This can imply a failure from a user's point view, but sometimes it is considered as an invalid failure (\eg{} requirements vs. usability issues) by developers. %imply a failure instead of imply on a failure 
In \emph{JabRef}, the undo/redo actions did not revert discrete operations. 
For example, pressing the button "Undo" clears all texts typed into different text fields instead of clearing only one field each time the button "undo" is pressed.

Another important point concerns the WIMP vs. post-WIMP GUIs faults.
We classified more failures involving WIMP than post-WIMP widgets.
A possible explanation is that, despite the increasing interactivity of GUIs, the analyzed GUIs still rely more on WIMP widgets and interactions.
Moreover, users now master the behavior of WIMP widgets so that they can easily identify when they provoke failures.
It may not be the case with \adhoc{} and post-WIMP widgets. %We discussed this concept in the Introduction

\begin{table*}[t]
\scriptsize
\caption{Mutants planted according to faults in the GUI Fault Model \label{table.mutsplant}}
\centering
\begin{tabular}{cccccccccccccccc}
\toprule
\textbf{ID} & GSA1 & GSA2 & GSA3 & DT1 & DT2 & DT3 & IB1 & ACT1& ACT2& ACT3 & RVSB1 & RVSB2 & RVSB3 & FDBK1 & FDBK2\\
\midrule
\textbf{\#Mutants} & 3 & 7 & 4 & 2 & 1 & 1 & 1 & 8 & 16 & 5 & 9 & 2 & - & 3 & 3 \\
\textbf{\#Length} & 0..2 & 0..4 & 0..4 & 4 & 2 & 4 & 4 & 1..7 & 1..8 & 4..5 & 2..8 & 2 & - & 2..5 & 3..4 \\
\midrule
%\multicolumn{2}{c}{- no mutant planted}
\end{tabular}
%\caption{Mutants planted according to faults in the GUI Fault Model \label{table.mutsplant}}
\end{table*}

\section{Are GUI Testing Tools Able to Detect Classified Failures? An Empirical Study}\label{sec.eval3}

This section provides an empirical study of two GUI testing tools: GUITAR \cite{NGU13} and Jubula\footnote{\url{http://www.eclipse.org/jubula}}. 
To demonstrate the current limitations of GUI testing tools in testing real GUIs,
we applied those tools to detect the failures previously classified into our GUI fault model.

\subsection{GUITAR and Jubula}

GUITAR is one of the most widespread academic GUI testing tools. 
It extracts the GUI structure by reverse engineering.
This structure is transformed into a GUI Event Flow Graph (EFG), where each node represents a widget event.
Based on this EFG, test cases are generated and executed automatically over the SUT.
We used the plugin for Java Swing (\ie{} JFC GUITAR version 1.1.1)\footnote{\url{http://sourceforge.net/apps/mediawiki/guitar/}}.
In GUITAR, each test case is composed by a sequence of widget events.
The generation of test cases can be parameterized with the size of that sequence (\ie{} test case length). 

Jubula is a semi-automated GUI testing tool that leverages pre-defined libraries to create test cases. 
These libraries contain modules that can be reused to generate manually test sequences.
%% [FIXED]FIXME What are actions? Give an example. Why/How it aims at covering WIMP/post-WIMP?
The modules encompass actions (\eg{} check, select) and interactions (\eg{} click, drag and drop) over different GUI toolkits (\eg{} swing, SWT, RCP, mobile).
%Moreover, some actions/interactions are available only for specific GUI toolkits.
%Testing figure properties on canvas only can be done for RCP SUT's.
%[FIXED]FIXME rephrase the following sentence. Not clear
We have reused the library dedicated to Java Swing (Jubula version 7.2) to write the test cases presented in the next experiments.
This library contains actions to test only standard widgets such as dragging a column/row of a table by passing an index. 
To test \adhoc{} widgets (\eg{} canvas), we made a workaround by mapping actions directly to these widgets.
For example, to draw a shape on canvas we need to specify the exact position (\eg{} drag and drop coordinates) where the interaction should be executed.

\subsection{Experiment}

We selected JabRef\footnote{\url{http://jabref.sourceforge.net/}}, a software to manage bibliographic references.
JabRef is written in Java which allows us to apply both GUITAR and Jubula.
For each fault described in our GUI fault model, we selected one reported failure.
To reproduce each failure, we downloaded the corresponding faulty version of JabRef. %or reintroduced the error using a patch or commit.
We used the exact test sequence (\ie{} number of actions) to reproduce a failure.
In GUITAR, all test cases were generated automatically over a faulty version.
In Jubula, each test case was created manually to detect one failure. %by reusing available libraries.
All test cases were written by one of the authors of this paper who has expertise in JabRef.
Also, their test sequences are extracted by analyzing failure reports (\eg{} steps to reproduce a failure) and reusing Jubula's libraries.
Then, GUITAR and Jubula run all their test cases automatically for checking whether the selected failure is found.

\subsection{Results and Discussion}

\begin{table}[th]
\scriptsize\setlength{\tabcolsep}{0.2cm}
\caption{JabRef failures detected by GUITAR and Jubula\label{table.failuresRes}}
\centering
\begin{tabular}{p{0.5cm}p{0.7cm}p{4cm}cc}
\toprule
\textbf{ID fault} & \textbf{ID failure} & \textbf{Bug repository link} & \textbf{GUITAR} & \textbf{Jubula} \\
\midrule
GSA1& \#1 &\url{http://sourceforge.net/p/jabref/bugs/160/} & \notok & \notok \\
GSA2& \#2 &\url{http://sourceforge.net/p/jabref/bugs/514/} & \notok & \ok  \\
GSA3& \#3 &\url{http://sourceforge.net/p/jabref/bugs/166/} & \notok & \ok \\
DT1  & \#4 &\url{http://sourceforge.net/p/jabref/bugs/716/} & \notok & \ok  \\
DT2  & \#5 & - &  &  \\
DT3  & \#6 &\url{http://sourceforge.net/p/jabref/bugs/575/}  & \notok & \ok \\
IB1    & \#7 & - &  & \\
ACT1& \#8 &\url{http://sourceforge.net/p/jabref/bugs/495/}  & \ok  & \ok \\
ACT2& \#9 &\url{http://sourceforge.net/p/jabref/bugs/536/}  & \ok  & \ok \\
ACT3& \#10&\url{http://sourceforge.net/p/jabref/bugs/809/}  & \notok &\notok  \\
RVSB1& \#11&\url{http://sourceforge.net/p/jabref/bugs/560/}  & \notok & \ok \\
RVSB2& \#12& -  &   &   \\
RVSB3& \#13&\url{http://sourceforge.net/p/jabref/bugs/458/}  & \ok  & \ok  \\
FDBK1& \#14&\url{http://sourceforge.net/p/jabref/bugs/52/}  & \notok  & \ok \\
FDBK2& \#15& -  &   &  \\
\bottomrule
\end{tabular}
%\caption{JabRef failures detected by GUITAR and Jubula\label{table.failuresRes}}
\end{table}

Table \ref{table.failuresRes} summarizes the detection of the JabRef GUI failures by GUITAR and Jubula.
These failures cover 11 out of the 15 faults described in our fault model. 
The remaining four faults were not covered for two reasons:
\begin{inparaenum}
\item no failure was classified for that fault; or
%\item a failure was classified, but we could not reproduce it, since it only occurred in a specific environment (\eg{} Operating System) or 
\item a failure was classified, but we could not reproduce it - only occurred in a specific environment (\eg{} Operating System) or 
given a certain input (\eg{} a particular database in JabRef).
\end{inparaenum}
%For example, some failures only manifest in a specific environment (\eg{} Operating System) or 
%they only occur given a certain input entry (\eg{} a particular database in JabRef).
%Thus, although we classified failures in our fault model, which required their root cause,
%we could not reproduce some of them to run our experiment.

%these failures are hard to reproduce, even though we can classify them in our fault model. %which only needs the root cause.

The reported failures in JabRef are mostly related to WIMP widgets, so we would expect GUITAR and Jubula to detect them,
but it was not the case.
%Yet, some of them were not detected by GUITAR and Jubula.
For instance, failure \#1 reports an incorrect display of buttons' label;
its root cause is the incorrect size of a widget positioned to the left of them.
%However, its root cause is the incorrect size of a widget positioned to the left of them. %another widget (\ie{} a panel widget positioned to the left of them).
Thus, this failure does not affect the values of internal properties (\eg{} text, event handlers) of those buttons. %internal properties values
%%[FIXED]FIXME the following sentence is not clear.
%In GUITAR, checking the panel properties did not detect this failure since the value of size property (\eg{} width, height) remained the same.
In GUITAR, checking the properties of that widget did not reveal this failure since the expected and actual values of its size property (\eg{} width) remained the same. 
In Jubula, the concerned widget cannot be mapped to test cases execution and thus cannot be tested.

Failures \#2 and \#3 refer to an incorrect menu path and a misspelling, respectively. 
%%[FIXED]FIXME the following sentence is not clear.
%Both failures were detected by Jubula since the expected values of widgets properties are defined manually.
Both failures were detected by Jubula.
However, these failures were not found by GUITAR.
Indeed, GUITAR does reverse engineering of an existing GUI to produce tests.
If this GUI is faulty, GUITAR will produce tests that will consider these failures as the correct behavior.

Failures \#8 and \#13 that lead to a crash of the GUI were found by both GUITAR and Jubula.
However, failures \#4, \#6, \#10, and \#11 that affect the data model were not detected by GUITAR for two reasons.
%% FIXME Explain where does this table entry come from.
First, GUITAR does not test the table entries in JabRef since they represent the data model.
To do this, we need to extend GUITAR to interact with them.
%% [FIXED]The following sentence is not clear.
%Second, the events fire properly (\eg{} no exception) and GUI properties are the "expected" ones.
Second, the test cases successfully passed, but a failure has been revealed.
That means, the events are fired properly (\eg{} no exception) and GUI properties are the "expected" ones.
For example, a text property of a status bar contains the value: \emph{"Redo: change field"}, when this action was actually not redone.
Similarly, failure \#10 was not detected by Jubula.
This failure reports an unexpected auto-completion when the action "save" is triggered by shortcuts.
We reproduced this failure manually but the test case was successfully replayed by Jubula.
The input text via keyboard was typed and saved automatically without any interference of the auto-completion feature.

Another point is the accuracy of test cases generated manually in Jubula.
%When the test cases are created manually testers are free to write them.
Detecting failure \#6 depends on how the test case is written.
For example, adding a field that contains LaTeX commands (\eg{} 100$\backslash$\%), and 
then checking its output in a preview window should not contain any command (\eg{} 100\%).  
So, we can write a test case to test the outputs in the preview window only looking for commands (\eg{} \emph{SelectPattern[\%, equals] in ComponentText[preview]}).
Or, write a test case to check whether an entire text matches to the expected one (\eg{} \emph{CheckText[100\%, equals] in ComponentText[preview]}).
However, the last test case will fail since a text from preview window in JabRef is shown internally as HTML 
and, in Jubula, the action's parameters cannot be specified in that format.
%and, in Jubula, the parameters defined for actions (\eg{} CheckText) cannot be specified in that format.
%Consequently, if we have knowledge about the failure we can select a more suitable action. 

Our experiment does not aim at comparing both tools since GUITAR is a fully automated tool contrary to Jubula.
%which requires a major effort to maintain and evolve test suites \cite{GRE09}.
However, the results of this study highlight the current limitations of GUI testing tools.
%%[FIXED]FIXME what does it mean "when their values are affected? Not clear
GUITAR and Jubula currently mainly work for detecting failures that affect properties of standard widgets.
Moreover, GUITAR does GUI regression testing:
it considers a given GUI as the reference one from which tests will be produced.
If this GUI is faulty, GUITAR will produce tests that will consider these failures as the correct behavior.
A possible solution to overcome this issue is to base the test process on the specifications (requirements, \emph{etc.}) of the GUI.

\section{Forging faulty GUIs for benchmarking}\label{sec.eval2}

In this section, we evaluate the usefulness of our fault model by applying it on a highly interactive open-source software system.
We created mutants of this system corresponding to the different faults of the model.
The main goal of these mutants is to provide GUI testers with benchmark tools to evaluate the ability of GUI testing tools to detect GUI failures.
As an illustration of the practical use of these mutants, we executed two GUI testing tools against the mutants of the system.
Thanks to that we caught a glimpse of their ability to cover our proposed fault model.
The goal of this experiment is to answer the research question: 
\emph{what are the benefits of this fault model for GUI testing?}

\subsection{Mutants Generation}
As highlighted by Zhu \etal{}, \emph{"software testing is often aimed at detecting faults in software. 
A way to measure how well this objective has been achieved is to plant some artificial faults into the program and check if they are detected by the test. 
A program with a planted fault is called a mutant of the original program"} \cite{ZHU97}.
Following this principle, we planted 65 faults in a highly interactive open-source software system, namely Latexdraw\footnote{\url{http://sourceforge.net/projects/latexdraw/}}, using our proposed fault model.
Latexdraw has been selected because of the following points:
\begin{inparaenum}
\item it is a highly interactive system written in Java and Scala (dedicated to the creation of drawings for \LaTeX);
\item its GUI mixes both standard and \adhoc{} widgets;
\item it is released under an open-source license (GPL2) so that it can be freely used by the testing community.
\end{inparaenum}

We created 65 mutants corresponding to the different faults of our proposed fault model.
All these mutants and the original version are freely available\footnote{\url{https://github.com/arnobl/latexdraw-mutants}\label{fn1}}.
Each mutant is documented to detail its planted fault and the oracle permitting to find it\footnoteref{fn1}. % \fixme{fault/error/failure}
Multiple mutants have been created from each fault by:
using WIMP (22 mutants) or post-WIMP (43 mutants) widgets to kill the mutants;
varying the test case length (\ie{} the number of actions required to provoke the failure).
Each action (\eg{} select a shape) requires a minimal number of events (\eg{} in LaTeXDraw a DnD requires at least three events: press/move/release) to be executed.
Table~\ref{table.mutsplant} summarizes the number of forged mutants and the minimal and maximal test case length for each fault.
For instance, a length $0..2$ means there exists at least one mutant requiring a minimum of 0 action or a maximum of two actions).
%However, the fault RVSB3 is currently not covered by the Latexdraw mutants as well as several WIMP faults (\eg{} IB1, DT1).
However, the fault RVSB3 is currently not covered by the Latexdraw mutants.
%Similarly, some mutants planted into fault categories only rely on post-WIMP interactions or widgets (\eg{} IB1, DT1).
Similarly, some planted mutants only rely on post-WIMP interactions or widgets (\eg{} IB1, DT1).

\subsection{How GUI testing tools kill our GUI mutants: a first experiment}

We applied the GUI testing tools GUITAR and Jubula on the mutants to evaluate their ability to kill them.
Our goal is not to provide benchmarks against these tools but rather highlight the current challenges in testing interactive systems not considered yet (\eg{} post-WIMP interactions). 
GUITAR test cases have been generated automatically while Jubula ones have been written manually.

Considering the mutants planted at the user interface level, Jubula and GUITAR tests killed the mutants that involve checking standard widget properties,
such as layout (\eg{} width, height) and state (\eg{} enable, selection, focusable). %, existence). 
Also, it is possible to test simple data (\eg{} string values on text fields) on those widgets.
However, most of the mutants that concern the \adhoc{} widgets were alive.
Notably, when test cases involve testing complex data from the data model.
For example, it is not possible to compare the results of the actual shape on canvas against the expected one. 
Even if some shape properties (\eg{} rotation angle) are presented on standard widgets (\eg{} spinner), 
GUITAR and Jubula cannot state whether the current values in these widgets match the expected shape rotation on the canvas.

Likewise, our GUITAR and Jubula tests cannot kill most of the user interaction mutants that result on a wrong presentation of shapes.
In particular, when we tested mutants planted into the \emph{Reversibility} or \emph{Feedback} categories.
For example, testing undo/redo operations in Latexdraw should compare all states to manipulate a shape on canvas.
Moreover, the tests verdict on Jubula passed even though interactions are defined incorrectly (\eg{} mouse cursor does not follow a DnD) or actions cannot be executed (\eg{} a button is disabled).
In GUITAR, the generated test cases do not cover properly actions having dependencies.
For example, the action "\emph{Delete}" in Latexdraw requires first selecting a shape on canvas.
However, no test sequence that contains  "\emph{Select Shape}" before "\emph{Delete Shape}" was generated. 
Thus, some mutants could not be killed.

Table \ref{table.mutresults} gives an overview of the number of mutants killed by GUITAR and Jubula.
The results show that both tools are not able to kill all mutants because of the four following reasons:
\begin{inparaenum}
\item \emph{Testing Latexdraw with GUITAR and Jubula is limited to the test of the standard Swing widgets}.
In Jubula, the test cases are only written according to libraries available in the Swing toolkit.
In GUITAR, the basic package for Java Swing GUIs only covers standard widgets and mono-events (\eg{} a click on a button).
\item \emph{Configure or customize a GUI testing tool to test post-WIMP widgets is not a trivial task}.
For example, each sequence of a test case in Jubula needs to be mapped for the corresponding GUI widget manually.
Also, GUITAR needs to be extended to generate test cases for \adhoc{} widgets (\eg{} canvas) as well their interactions (\eg{} multi-modal interactions).
\item \emph{Testing post-WIMP widgets requires a long test case sequence}.
In Latexdraw, a sequence to test interactions over these widgets is composed of at least two actions. 
That sequence is longer when we have to detect failures into undo/redo operations.
\item \emph{It is not possible to give a test verdict for complex data}.
The oracle provided by the two GUI testing tools do not know the internal behavior of \adhoc{} widgets, their interaction features and data presentation.
\end{inparaenum}
%These first experiments highlight the benefits of our fault model for measuring the ability of GUI testing tools in finding GUI failures. % Should we answer directly the RQ?
These results answer the research question by highlighting the benefits of our fault model for measuring the ability of GUI testing tools in finding GUI failures.

\begin{table}[h]
\scriptsize
\caption{Mutants killed by GUITAR and Jubula \label{table.mutresults}}
\centering
\begin{tabular}{p{0.6cm}cccc}
\toprule
 & \multicolumn{2}{c}{\textbf{GUITAR}} & \multicolumn{2}{c}{\textbf{JUBULA}} \\
\midrule
\textbf{ID} & WIMP & post-WIMP & WIMP & post-WIMP \\
GSA1& 2 & 0 & 2 & 0 \\
GSA2&  5 & 0 & 6 & 1 \\
GSA3&  3 & 0 & 3 & 0 \\
DT1  &  - & 0 & - & 0 \\
DT2  &  - & 0 & - & 0 \\
DT3  &  - & 0 & - & 1 \\
IB1    &  - & 0 & - & 0 \\
ACT1&  0 & 0  & 0 & 1 \\
ACT2&  3 & 0  & 3 & 0 \\
ACT3&  2 & 0  & 2 & 0 \\
RVSB1&  2 & 0  & 2 & 0 \\
RVSB2& - & 0  & - & 0 \\
RVSB3& - & -  & - & - \\
FDBK1& 1 & 0  & 1 & 0 \\
FDBK2& - & 0  & - & 0 \\
\midrule
%\multicolumn{2}{c}{- no mutant planted}
\end{tabular}
%\caption{Mutants killed by GUITAR and Jubula \label{table.mutresults}}
\end{table}

\subsection{Threats to Validity}

Regarding the conducted empirical studies, we identified the two following threats to validity.
The first one concerns the scope of the proposed fault model since we evaluated it empirically on a small number (five) of interactive systems.
To limit this threat, we selected interactive systems that cover different aspects of the HCI concepts we detailed in Section~\ref{sec.back}.
The second threat we identified concerns the subjectivity observed in bug reports to describe failures.
To deal with this, we based the classification on the bug report artifacts (patches, logs, \etc) to identify the root cause of the reported failures.

%The first one concerns if the fault model is \emph{generic} enough.
%The goal of the proposed fault model is to describe faults at an abstract level. % to classify them into categories relevant in a testing purpose. 
%They are classified into macro fault categories for a GUI testing purpose.
%We chose a certain level of detail/abstraction that, for instance, does not permit to classify into a specific domain 
%such as Virtual Reality.
%However, we can reuse this fault model to help us to identify failures for a specific domain.
%For example, multimodal interfaces bring different types of fusion. 
%Missing a good example 
%So, fusion failures are classified into IB1 "incorrect behavior" of a user interaction level.

%The second threat is the \emph{reliability} of the proposed fault model.
%We applied our fault model on 5 applications.
%The goal of this experiment was not to demonstrate that our fault model covers the most application domains but rather
%the fault model is relevant to classify GUI failures that come from different type of interactions, widgets, and also platforms.

%The last treat concerns that one author categorized all failures in the fault model.
%It is not easy to identify and classify failures from a black box analysis perspective. 
%To deal with this, we selected bug reports that contains more complete descriptions and artifacts (patches, execution trace, \emph{etc})...}
%\fi

\section{Related Work}\label{sec.related}

Existing fault classifications are presented in a higher level of abstraction
considering mainly the components that are affected by faults.
Most classifications leverage the software assets (\eg{} specification, models, architecture, code) to define their faults.
These faults have been described into fault models \cite{BOC91, PRE13} or defects taxonomies \cite{RAM92}.

In an effort to cover GUIs, the Orthogonal Defect Classification (ODC) \cite{RAM92} is extended by IBM Research to include GUIs faults.
These faults focus on the appearance of widgets, navigation between widgets, 
unexpected behavior of widgets events and input devices.
In our fault model, we do not cover faults that concern the behavior of input devices (\ie{} hardware fault). 
Although this taxonomy considers GUI faults, it does not separate the user interface and user interaction faults. 
Moreover, this extension does not consider faults caused by post-WIMP widgets and their advanced interactions as well faults into the \emph{data presentation} category.

Li \etal{} categorize faults of industrial and open source projects using the ODC taxonomy \cite{NIN10}.
The category \emph{Interface} concerns several GUI defects.
However, this single category covers several user interface defects related to specific widgets such as \emph{window}, \emph{title bar}, \emph{menu}, or \emph{tool bar}.
Similarly, the interaction defects are limited to \emph{mouse} and \emph{keyboard}.
Thus, it is not possible to identify the kind of faults classified into these categories since they are not detailed.
For example, a fault classified into the \emph{mouse} category can concern an interaction, an action, or an input device. 

Brooks \etal{} \cite{BRO09} present a study that characterizes GUIs based on reported faults of three industrial systems.
%% FIXME the following sentence is not clear. "by including other categories such as GUI defects"
To classify all these faults (GUI and non-GUI faults), the authors adapted a defect taxonomy by including other categories such as GUI defects.
This category encompasses both the user interface and user interaction faults. 
%This category covers the user interface and user interaction.
Also, {B\o rretzen} \etal{} \cite{JON06} analyze faults reported by four projects by combining two defect taxonomies.
Both works introduce a category that concerns the GUI faults but these faults are not described and thus no classification is presented.
Strecker \etal{} \cite{STR08} characterize faults that affect GUI test suites.
However, these faults do not concern the GUI faults but any fault at the code level (\eg{} class or method faults) that may affect the GUI. 

%% FIXME the following sentence is not clear. "GUI failures instead of GUI faults" What does it mean?
In contrast, several research papers concern the fault effects by classifying GUI failures instead of GUI faults.
In general, these works focus on specific GUIs (automotive GUIs \cite{MAU12}) or domains (mobile \cite {KUM10}, safety-critical \cite{LUT04}).
For example, Maji \etal{} characterize failures for mobile operating systems \cite {KUM10}. 
These failures are classified according to the fault localization.
For example, a failure manifested in a \emph{camera} is categorized in the \emph{Camera segment}. 
Similarly, failures for other segments such as Web, Multimedia, or GUI are categorized. 
Also, Zaeem \etal{} \cite {ZAE14} have conducted a bug study for Android applications to automate oracles.
They identified 20 categories including some GUI issues such as Rotation (device's rotation), Gestures (zooming and out) and Widget. 
Although, these papers have investigated failures in a context that brings many advances in terms of interactive features, no classification or discussion about these kinds of failures is presented.

Mauser \etal{} propose a GUI failure classification for automotive systems \cite{MAU12}. 
This classification is based on the three categories:
design, content, and behavior.
In the \emph{Design} category, the failures refer to GUI layouts (\eg{} color, font, position). 
In the \emph{Content} category, the failures are associated to data displayed such as text, animation, and symbols/icons. 
The failures in the \emph{Behavior} category are caused by a wrong behavior of windows (\eg{} wrong pop-up) or widgets (\eg{} wrong focus). 
The authors focus on characterizing GUI failures based only on a small set of specific widgets designed for these kinds of GUIs. 
Furthermore, they do not consider user interaction failures.

\section{Conclusion and Research Agenda}\label{sec.conclu}

This paper proposes a GUI fault model for providing GUI testers with benchmark tools to evaluate the ability of GUI testing tools to detect GUI failures.
This fault model has been empirically assessed by analyzing and classifying into it 279 GUI bug reports of different open-source GUIs.
To demonstrate the benefits of the proposed fault model, mutants have then been developed from it on a Java open-source GUI. 
As an illustrative use case of these mutants, we executed two GUI testing tools on these mutants to evaluate their ability to detect them.
This experiment shows that if current GUI testing tools have demonstrated their ability for finding several kinds of GUI errors, they also fail at detecting several GUI faults we identified.
The underlying reasons are twofold.
First, GUI failures may be related to the graphical rendering of GUIs.
Testing a GUI rendering is a complex task since current testing techniques mainly rely on code analysis that can hardly capture graphical properties.
Second, the current trend in GUI design is the shift from designing GUIs composed of standard widgets to designing GUIs relying on more complex interactions and \adhoc{} widgets~\cite{BEA00b,BEL04,BLO10}.
New GUI testing techniques have thus to be proposed for fully testing, \emph{as automated as possible}, GUI rendering and complex %interactions and widgets.
interactions using \adhoc{} widgets.

\section*{Acknowledgements}
This work is partially supported by the French BGLE Project CONNEXION.

\balance
\bibliographystyle{IEEEtran}
\bibliography{ref}

\begin{thebibliography}{10}
\providecommand{\url}[1]{#1}
\csname url@samestyle\endcsname
\providecommand{\newblock}{\relax}
\providecommand{\bibinfo}[2]{#2}
\providecommand{\BIBentrySTDinterwordspacing}{\spaceskip=0pt\relax}
\providecommand{\BIBentryALTinterwordstretchfactor}{4}
\providecommand{\BIBentryALTinterwordspacing}{\spaceskip=\fontdimen2\font plus
\BIBentryALTinterwordstretchfactor\fontdimen3\font minus
  \fontdimen4\font\relax}
\providecommand{\BIBforeignlanguage}[2]{{%
\expandafter\ifx\csname l@#1\endcsname\relax
\typeout{** WARNING: IEEEtran.bst: No hyphenation pattern has been}%
\typeout{** loaded for the language `#1'. Using the pattern for}%
\typeout{** the default language instead.}%
\else
\language=\csname l@#1\endcsname
\fi
#2}}
\providecommand{\BIBdecl}{\relax}
\BIBdecl

\bibitem{GAM95}
E.~Gamma, R.~Helm, R.~Johnson, and J.~Vlissides, \emph{Design patterns:
  elements of reusable object-oriented software}.\hskip 1em plus 0.5em minus
  0.4em\relax Addison-Wesley, 1995.

\bibitem{BEA00b}
M.~Beaudouin-Lafon, ``Instrumental interaction: an interaction model for
  designing post-{WIMP} user interfaces,'' in \emph{Proc. of CHI'00}.\hskip 1em
  plus 0.5em minus 0.4em\relax ACM, 2000, pp. 446--453.

\bibitem{MEM07}
A.~M. Memon, ``An event-flow model of {GUI}-based applications for testing,''
  \emph{STVR}, vol.~17, no.~3, pp. 137--157, 2007.

\bibitem{COH12}
M.~Cohen, S.~Huang, and A.~Memon, ``Autoinspec: Using missing test coverage to
  improve specifications in {GUIs},'' in \emph{Proc of ISSRE'12}, 2012, pp.
  251--260.

\bibitem{ARL12}
S.~Arlt, A.~Podelski, C.~Bertolini, M.~Schaf, I.~Banerjee, and A.~Memon,
  ``Lightweight static analysis for {GUI} testing,'' in \emph{Proc of
  ISSRE'12}, 2012.

\bibitem{MAR12}
L.~Mariani, M.~Pezz\`{e}, O.~Riganelli, and M.~Santoro, ``Autoblacktest:
  Automatic black-box testing of interactive applications,'' in \emph{Proc. of
  ICST'12}.\hskip 1em plus 0.5em minus 0.4em\relax IEEE, 2012, pp. 81--90.

\bibitem{NGU10}
D.~H. Nguyen, P.~Strooper, and J.~G. S\"{u}\ss, ``Automated functionality
  testing through {GUIs},'' in \emph{Proc. of ACSC '10}, 2010, pp. 153--162.

\bibitem{BEL04}
M.~Beaudouin-Lafon, ``Designing interaction, not interfaces,'' in \emph{Proc.
  of AVI'04}, 2004.

\bibitem{BLO10}
A.~Blouin and O.~Beaudoux, ``Improving modularity and usability of interactive
  systems with {Malai},'' in \emph{Proc. of EICS'10}, 2010, pp. 115--124.

\bibitem{DAM97}
A.~van Dam, ``Post-{WIMP} user interfaces,'' \emph{Commun. ACM}, vol.~40,
  no.~2, pp. 63--67, Feb. 1997.

\bibitem{BOC91}
G.~von Bochmann, A.~Das, R.~Dssouli, M.~Dubuc, A.~Ghedamsi, and G.~Luo, ``Fault
  models in testing.'' in \emph{Protocol Test Systems}, 1991, pp. 17--30.

\bibitem{SHN83}
B.~Shneiderman, ``Direct manipulation: a step beyond programming languages,''
  \emph{IEEE Computer}, vol.~16, no.~8, pp. 57--69, 1983.

\bibitem{HUT85}
E.~L. Hutchins, J.~D. Hollan, and D.~A. Norman, ``Direct manipulation
  interfaces,'' \emph{Hum.-Comput. Interact.}, vol.~1, no.~4, pp. 311--338,
  1985.

\bibitem{NOR02}
D.~A. Norman, \emph{The Design of Everyday Things}, reprint paperback~ed.\hskip
  1em plus 0.5em minus 0.4em\relax Basic Books, 2002.

\bibitem{APP12}
C.~Appert, O.~Chapuis, and E.~Pietriga, ``Dwell-and-spring: undo for direct
  manipulation,'' in \emph{Proc. of CHI'12}.\hskip 1em plus 0.5em minus
  0.4em\relax ACM, 2012, pp. 1957--1966.

\bibitem{PRE13}
A.~Pretschner, D.~Holling, R.~Eschbach, and M.~Gemmar, ``A generic fault model
  for quality assurance,'' in \emph{Proc of MODELS'13}, 2013.

\bibitem{BLO11}
A.~Blouin, B.~Morin, G.~Nain, O.~Beaudoux, P.~Albers, and J.-M. J\'ez\'equel,
  ``Combining aspect-oriented modeling with property-based reasoning to improve
  user interface adaptation,'' in \emph{Proc. of EICS'11}, 2011, pp. 85--94.

\bibitem{APP08}
C.~Appert and M.~Beaudouin-Lafon, ``{SwingStates: Adding state machines to Java
  and the Swing toolkit},'' \emph{SW: Practice and Experience}, vol.~38,
  no.~11, pp. 1149--1182, 2008.

\bibitem{NGU13}
B.~Nguyen, B.~Robbins, I.~Banerjee, and A.~Memon, ``{GUITAR}: an innovative
  tool for automated testing of {GUI}-driven software,'' \emph{Automated
  Software Engineering}, pp. 1--41, 2013.

\bibitem{ZHU97}
H.~Zhu, P.~A.~V. Hall, and J.~H.~R. May, ``Software unit test coverage and
  adequacy,'' \emph{ACM Comput. Surv.}, vol.~29, no.~4, pp. 366--427, 1997.

\bibitem{RAM92}
R.~Chillarege, I.~S. Bhandari, J.~K. Chaar, M.~J. Halliday, D.~S. Moebus, B.~K.
  Ray, and M.-Y. Wong, ``Orthogonal defect classification-a concept for
  in-process measurements,'' \emph{IEEE Trans. Softw. Eng.}, vol.~18, no.~11,
  pp. 943--956, 1992.

\bibitem{NIN10}
N.~Li, Z.~Li, and X.~Sun, ``Classification of software defect detected by
  black-box testing: An empirical study,'' in \emph{Proc. of WCSE'10}.

\bibitem{BRO09}
P.~Brooks, B.~Robinson, and A.~Memon, ``An initial characterization of
  industrial graphical user interface systems,'' in \emph{Proc. of ICST'09}.

\bibitem{JON06}
J.~A. B{\o}rretzen and R.~Conradi, ``Results and experiences from an empirical
  study of fault reports in industrial projects,'' in \emph{Proc. of
  PROFES'06}.\hskip 1em plus 0.5em minus 0.4em\relax Berlin, Heidelberg:
  Springer-Verlag, 2006, pp. 389--394.

\bibitem{STR08}
J.~Strecker and A.~Memon, ``Relationships between test suites, faults, and
  fault detection in gui testing,'' in \emph{Proc. of ICST'08}, 2008, pp.
  12--21.

\bibitem{MAU12}
D.~Mauser, A.~Klaus, R.~Zhang, and L.~Duan, ``{GUI} failure analysis and
  classification for the development of in-vehicle infotainment,'' in
  \emph{Proc. of VALID'12}, 2012, pp. 79--84.

\bibitem{KUM10}
A.~Kumar~Maji, K.~Hao, S.~Sultana, and S.~Bagchi, ``Characterizing failures in
  mobile oses: A case study with android and symbian,'' in \emph{Proc. of
  ISSRE'10}, 2010, pp. 249--258.

\bibitem{LUT04}
R.~Lutz and I.~mikulski, ``Empirical analysis of safety-critical anomalies
  during operations,'' \emph{IEEE Trans. Softw. Eng.}, pp. 172--180, 2004.

\bibitem{ZAE14}
R.~N. Zaeem, M.~R. Prasad, and S.~Khurshid, ``Automated generation of oracles
  for testing user-interaction features of mobile apps,'' in \emph{Proc. of
  ICST'14}, 2014.

\end{thebibliography}

\end{document}